\documentclass{article}

\usepackage[utf8]{inputenc}
\usepackage[T1]{fontenc}
\usepackage[english]{babel}
\usepackage{amsmath,amssymb,amsfonts}
\usepackage{graphicx}
\usepackage{booktabs}
\usepackage{array}
\usepackage{multirow}
\usepackage{hyperref}
\usepackage[numbers]{natbib}
\usepackage{ragged2e}
\usepackage{geometry}
\usepackage{enumitem}
\usepackage{float}
\usepackage{caption}
\usepackage{subcaption}
\usepackage{indentfirst}

\addto\captionsenglish{}
\DeclareCaptionFont{sevenpt}{\fontsize{7}{8}\selectfont}
\hypersetup{hidelinks}
\newcolumntype{L}[1]{>{\raggedright\arraybackslash}p{#1}}
\newcolumntype{C}[1]{>{\centering\arraybackslash}p{#1}}
\newcolumntype{R}[1]{>{\raggedleft\arraybackslash}p{#1}}

\title{\textbf{Information Networks of Stock Prices}}

\author{%
  Muhammad Aldy Hassan\thanks{Scholar in Dept. Computational Sociology, Bandung Fe Institute, \texttt{aldy@compsoc.bandungfe.net}}
  \and
  Hokky Situngkir\thanks{AI Center, Institut Teknologi Del, \texttt{hokky.situngkir@del.ac.id}}
}
\date{}

\begin{document}

\maketitle

\begin{abstract}
\justifying
The collective movement of stock prices harbors complex interdependencies that are conventionally simplified only through a linear lens. This paper explores computed structural network representations in the Indonesian capital market by testing the limits of Pearson correlation and Mutual Information (MI) in unveiling the spectral dynamics of the market. Across 2,328 rolling observation windows from 2015 to 2025, we examine 24 methodological configurations that combine three dependency estimators (Pearson, MI adaptive binning, and MI-kNN), two graph filtering schemes (Minimum Spanning Tree/MST and Planar Maximally Filtered Graph/PMFG), and four community decoders.

The empirical results unveil a fundamental reality: topological richness does not always resonate with sectoral classification precision. The Pearson, MST, and Infomap configuration is shown to remain the most robust foundation for recovering conventional sectoral taxonomy. Nevertheless, when deeper observation demands the exposition of local structures and the weave of heterogeneous communities, the architectural relaxation through PMFG demonstrates its superiority. In the realm of residual information detection, MI adaptive binning appears far more proportional than kNN; histogram-based regularization successfully tames empirical noise without sweeping away traces of non-linear dependency. Ultimately, the synergy of MI and PMFG is not positioned to dethrone the dominance of linear correlation, but rather to provide an essential analytical lens for excavating hidden economic sub-structures---such as the cohesion of commodity regimes---that have long transcended the rigid boundaries of the market's formal sectors.

\vspace{0.3cm}
\end{abstract}
\newpage

\section{Introduction}

Stock prices on the exchange constantly move within complex interdependencies driven by the cognitive collectivity of investors. The financial network approach is employed to simplify the spectrum of these complex dependency matrices in depicting such collectivity. Structural network construction frequently employs the \textit{Minimum Spanning Tree} (MST) approach to obtain the most efficient \textit{backbone} of inter-stock relationships. On the other hand, there exists the \textit{Planar Maximally Filtered Graph} (PMFG) approach, which preserves the same \textit{backbone} while allowing the formation of more complex local structures as long as planarity is satisfied \cite{ref1} \cite{ref2b}, \cite{ref11}, \cite{ref18}-\cite{ref24}. The success of these methods is generally measured by the extent to which the resulting communities align with economic groupings, ranging from industrial sectors and business groups to broader market hierarchies that depict the spectral collectivity of price movements on the trading floor \cite{ref21}, \cite{ref22}.

The most commonly used approach in relational calculations between stock price movements is the Pearson correlation. This approach is statistically stable in detecting linear co-movements. However, by definition, Pearson correlation is not designed to capture non-linear dependencies, \textit{tail dependence}, or \textit{non-Gaussian} relationships that frequently arise in financial return dynamics \cite{ref4}. Efforts to sharpen the correlative patterns provided by the Pearson correlation approach include, among others, the transfer entropy approach, which produces directed network diagrams that can be viewed as a form of macro causality between price movements \cite{ref2}. On the other hand, an alternative refinement using \textit{Mutual Information} (MI), rooted in Shannon's information theory, can be used to capture more local spectral dynamics as the complex informational relations contained within each stock price time series. MI is only equal to zero if the variables are truly independent, so theoretically it can detect both linear and non-linear relationships simultaneously \cite{ref9}, \cite{ref16}, \cite{ref18}.

The MI approach depends heavily on its estimation method. The use of \textit{adaptive binning} and k-Nearest Neighbors (kNN) reflects two opposing computational philosophies: regularization through \textit{coarse-graining} versus local density estimation \cite{ref4}, \cite{ref14}-\cite{ref16}. With respect to the use of PMFG, several studies on financial networks have explored graph filters using only static or single community detection algorithms. We demonstrate the potential of alternative detection algorithms such as \textit{greedy modularity}, Louvain, Leiden, and Infomap, which can produce highly varied levels of community detail \cite{ref6}, \cite{ref10}, \cite{ref12}-\cite{ref13}.

These approaches are relevant for application to exchanges in developing countries such as Indonesia. Factors such as sectoral structure, exposure to commodities, and liquidity imbalances render inter-stock relationships too complex to be simply reduced to ordinary linear correlation. Therefore, an explicit framework is constructed here to compare MI, PMFG, and particularly the MI-PMFG combination. At the outset, we explore and review the theoretical foundations employed, ranging from the construction of correlation and its informational relational model to the use of several community detection algorithms for deepening the groupings produced by the network graph model: \textit{greedy modularity}, Louvain, Leiden, and Infomap.

\section{Construction of Price Networks}

\subsection{Correlation networks, MST, and PMFG as economic taxonomy}

Correlation networks constitute a concise way to read the hierarchical structure of the market \cite{ref18}, \cite{ref19}. In this construction, the correlation $\rho_{ij}$ is usually mapped to a metric distance $d_{ij}=\sqrt{2(1-\rho_{ij})}$, after which MST selects N-1 edges that connect all N stocks without cycles and with minimum total distance \cite{ref1}, \cite{ref16}. By preserving the \textit{backbone} of the resulting correlative network, MST produces a highly concise graph, making it easier to visualize and effective in mapping global hierarchy. However, as a consequence of this simplification, MST deliberately ignores many local structures that potentially still hold important economic information \cite{ref11}.

As a topological relaxation of this concept, the PMFG method offers an alternative. Starting from a weighted complete graph, every pair of stocks is sorted based on similarity level from the highest to the lowest. A new edge is then added only if the resulting graph maintains its planar property \cite{ref3}, \cite{ref20}, \cite{ref24}. Referring to Euler's bound for the number of nodes $N\ge 3$, a maximally planar graph is constrained to have at most $3N-6$ edges. This characteristic keeps PMFG relatively sparse, yet endowed with a structure far richer than the basic tree structure consisting of only $N-1$ edges. In the context of information filtering networks, PMFG is regarded as a representation that not only preserves the MST \textit{backbone} but also accommodates additional local relations satisfying the planarity test \cite{ref11}.

One major impact of the planarity constraint is that PMFG allows the formation of loops and small cliques while still constraining its local complexity to a readable level. At genus zero, the relevant clique forms are limited to 3-cliques and 4-cliques. This is where PMFG becomes more expressive than MST: sectoral cohesion, central nodes, and interlocking market blocks can emerge without forcing us back into the complexity of a complete graph. Various financial market studies indicate that these features are important when researchers wish to read the thickness of local structure, changes in centrality, and hierarchical shifts across time windows \cite{ref21}-\cite{ref23}.

\subsection{From Shannon Entropy to MI}

MI is most easily understood when derived directly from Shannon entropy \cite{ref17}. In the discrete case, the Shannon entropy of a random variable X with probability p(x) is the average \textit{self-information} -log p(x). The more dispersed the distribution, the greater the average uncertainty. In this formulation, the natural logarithm is used, so the unit is nats; if base 2 is used, the unit changes to bits \cite{ref8}, \cite{ref16}.

\begin{equation}
H(X)=-\sum_x p(x)\log p(x)
\end{equation}

Joint entropy measures the joint uncertainty of the pair (X,Y), while conditional entropy measures the residual uncertainty of X after Y is known. For two discrete variables, joint entropy and conditional entropy are written as follows.

\begin{equation}
H(X,Y)=-\sum_{x,y} p(x,y)\log p(x,y)
\end{equation}

\begin{equation}
H(X\mid Y)=-\sum_y p(y)\sum_x p(x\mid y)\log p(x\mid y)
\end{equation}

Since $p(x,y)=p(y)\,p(x\mid y)$, the \textit{chain rule} $H(X,Y)=H(Y)+H(X\mid Y)=H(X)+H(Y\mid X)$ applies. From this identity, MI can be written as a direct derivation from uncertainty reduction:

\begin{equation}
I(X;Y)=H(X)-H(X\mid Y)=H(Y)-H(Y\mid X)
\end{equation}

Substituting $H(X\mid Y)=H(X,Y)-H(Y)$ yields the symmetric form more frequently used in information theory:

\begin{equation}
I(X;Y)=H(X)+H(Y)-H(X,Y)
\end{equation}

MI can also be viewed as the \textit{Kullback-Leibler divergence} between the joint distribution and the product of its marginals:

\begin{equation}
I(X;Y)=H(X)-H(X\mid Y)
\end{equation}

\begin{equation}
I(X;Y)=-\sum_x p(x)\log p(x)+\sum_{x,y}p(x,y)\log p(x\mid y)
\end{equation}

\begin{equation}
I(X;Y)=\sum_{x,y} p(x,y)\log\frac{p(x\mid y)}{p(x)}
\end{equation}

\begin{equation}
I(X;Y)=\sum_{x,y} p(x,y)\log\frac{p(x,y)}{p(x)p(y)}
\end{equation}

This last form makes the zero point of MI explicit: $I(X;Y)=0$ if and only if $p(x,y)=p(x)p(y)$, that is, X and Y are independent. Thus, zero correlation does not necessarily imply the absence of dependency, but zero MI indeed signifies no shared information whatsoever.

For stock returns as continuous variables, Shannon entropy is extended to differential entropy. In continuous density notation, the form becomes

\begin{equation}
h(X)=-\int f_X(x)\log f_X(x)\,dx
\end{equation}

Unlike discrete entropy, differential entropy can take negative values and is sensitive to changes in scale, so the two cannot be interpreted in exactly the same way. Even so, the combined formulation $h(X)+h(Y)-h(X,Y)$ within MI retains invariance under smooth one-to-one transformations. Therefore, the continuous MI measurement remains valid and relevant as a benchmark of dependency between variables, even though each marginal entropy does not always carry a simple absolute meaning that can be interpreted directly \cite{ref8}, \cite{ref16}.

The relationship with correlation becomes very clear at the Gaussian \textit{baseline}. Suppose (X,Y) follows a bivariate Gaussian with covariance

\begin{equation}
\Sigma=\begin{bmatrix}\sigma_X^2 & \rho\sigma_X\sigma_Y\\ \rho\sigma_X\sigma_Y & \sigma_Y^2\end{bmatrix}
\end{equation}

For a d-dimensional Gaussian variable with covariance $\Sigma$, the differential entropy satisfies $h_G(Z)=\frac{1}{2}\log\!\left((2\pi e)^d\det\Sigma\right)$. Hence, for the bivariate case above, we obtain

\begin{equation}
h_G(X)=\frac{1}{2}\log\!\left(2\pi e\,\sigma_X^2\right)
\end{equation}

\begin{equation}
h_G(Y)=\frac{1}{2}\log\!\left(2\pi e\,\sigma_Y^2\right)
\end{equation}

\begin{equation}
h_G(X,Y)=\frac{1}{2}\log\!\left((2\pi e)^2\sigma_X^2\sigma_Y^2(1-\rho^2)\right)
\end{equation}

Substituting into the MI identity yields the cancellation of all scale components $\sigma_X$ and $\sigma_Y$, so that the pure Gaussian MI is determined solely by $\rho$:

\begin{equation}
I_G(\rho)=h_G(X)+h_G(Y)-h_G(X,Y)=-\frac{1}{2}\log(1-\rho^2)
\end{equation}

This is why MI in the \textit{Gaussian-linear} case is merely a monotone transformation of correlation. If all dependencies are truly Gaussian, MI provides no new information beyond what is already captured by $\rho$.

We can define the residual information as the difference between the empirical MI and the \textit{Gaussian-equivalent} \textit{baseline}:

\begin{equation}
\Delta I_{ij}=I_{\mathrm{emp}}(X_i,X_j)-I_G(\rho_{ij})
\end{equation}

Thus, $\Delta I\approx 0$ means that the dependency of stock pair i,j is essentially adequately explained by the \textit{Gaussian-linear} structure. Conversely, persistent $\Delta I>0$ signals the presence of tail co-movement components, \textit{switching regime}, asymmetry, or other \textit{non-Gaussianity} that remains after the linear-equivalent component has been removed. $\Delta I$ thus becomes an important variable in the effort to distinguish between what is read correlatively and what only emerges through MI.

\subsection{MI Estimation: Adaptive Binning}

The introduction of \textit{adaptive binning} stems from the fact that continuous MI values cannot simply be extracted from their theoretical formulation. While the mathematical formulation of MI demands precision from the full joint distribution, market dynamics offer only limited empirical return series that are noisy, heavy-tailed, and asymmetrically distributed. It is here that \textit{adaptive binning} paves the way to tame the wildness of such data. The observation series is mapped into quantile chambers, ensuring that both the data swirl at the center of the distribution and the anomalies in the tails are proportionally represented in the calculation. Through this logic, \textit{adaptive binning} transcends a mere mechanical computational procedure and transforms into an essential regularization instrument. This approach preserves the purity of the dependency signal from being distorted by empirical bias, dampening spurious fluctuations that often arise from histogram cells that are either overcrowded or left empty.

The subsequent computational challenge is rooted in the fact that, in continuous data processing, the density functions $f_X$, $f_Y$, and $f_{XY}$ are generally not explicitly known. One of the most practical approaches to address this problem is through the \textit{coarse-graining} technique. Continuous return data is first mapped into a number of bins, then analyzed using the Shannon formula based on the empirical distribution of the discretization. Let $B_X$ and $B_Y$ represent the bin indices for variables X and Y, $n_{uv}$ be the number of observations in cell (u,v), and $n_{u\bullet}$ and $n_{\bullet v}$ respectively denote the observation frequencies on the marginal bins of X and Y. Based on these definitions, the empirical probabilities can be formulated as follows:

\begin{equation}
p_{uv}=\frac{n_{uv}}{N},\quad p_{u\bullet}=\frac{n_{u\bullet}}{N},\quad p_{\bullet v}=\frac{n_{\bullet v}}{N}
\end{equation}

The Shannon plug-in estimator is then written as

\begin{equation}
H(B_X)=-\sum_u p_{u\bullet}\log p_{u\bullet}
\end{equation}

\begin{equation}
H(B_Y)=-\sum_v p_{\bullet v}\log p_{\bullet v}
\end{equation}

\begin{equation}
H(B_X,B_Y)=-\sum_{u,v} p_{uv}\log p_{uv}
\end{equation}

So that the MI Adaptive Binning follows directly from the entropy identity:

\begin{equation}
I_{AB}=H(B_X)+H(B_Y)-H(B_X,B_Y)
\end{equation}

\begin{equation}
I_{AB}=\sum_{u,v} p_{uv}\log\frac{p_{uv}}{p_{u\bullet}p_{\bullet v}}
\end{equation}

Adaptive binning is operationally defined as a quantile-adaptive or \textit{equal-frequency} histogram estimator. For each stock pair with N valid observations, the number of bins (B) is determined using $B=\max\!\left(3,\lfloor N^{1/3}\rfloor\right)$. The bin boundaries on each marginal distribution are then formed based on their quantile values, so that each marginal bin accommodates approximately $N/B$ observations. After that, the observation data is mapped into a joint contingency table, and the MI value is computed from its empirical frequencies. Essentially, this implementation is a Shannon \textit{plug-in estimator} regularized through the \textit{quantile binning} technique.

The fundamental strength of this approach lies in its ability to mitigate the sparsity of the data structure. Given the typically heavy-tailed dynamics of financial returns, the use of static bin intervals frequently gives rise to topological imbalance: leaving empty space in the distribution tails while observation mass becomes extremely concentrated at the center. The presence of \textit{quantile bins}-based partitioning successfully addresses this problem by distributing marginal density more proportionally. The impact is highly crucial: estimator variance can be suppressed, while stock pairs that genuinely possess ties are no longer immediately distorted into inflated MI values purely as a result of overly sparse histograms. As a result, the stability of inter-stock relational reading is more solidly established, even when using narrow observation windows \cite{ref14}, \cite{ref16}.

However, this mechanism invariably demands a methodological compromise. The consequence of compressing observations into bin compartments is the fading of part of the data's microscopic resolution. Overly local interaction dynamics, dependency traces hidden sporadically at the tails, and short-lived regime shifts are vulnerable to being swept away, drowned by the aggregation of the \textit{coarse-graining} process. Ultimately, \textit{adaptive binning} indeed offers a far more stable and straightforward foundation for mapping, but it is not always the most precise instrument for uncovering subtle and specific non-linear residuals.

\subsection{MI Estimation based on k-Nearest Neighbors (kNN): from Kozachenko--Leonenko to KSG}

As a methodological solution to the aggregate resolution limitations of \textit{adaptive binning}, the k-Nearest Neighbors (kNN) approach is integrated to dissect another side of the dependency spectrum. Fundamentally different from \textit{adaptive binning}, which translates the market through distribution compartments, kNN delves into collective dynamics by directly tracking the geometry of proximity from each observation point. The measurement paradigm shifts sharply: the analytical focus is no longer directed at the accumulation of data mass contained within a static bin, but rather at how closely the return fluctuations of a stock pair resonate within their \textit{local neighborhood} \cite{ref4}. Through this architectural philosophy, kNN serves an essential function as a high-resolution and far more reactive analytical comparator. This approach is deliberately designed to accommodate and capture traces of non-linear relations.

The kNN estimation initiates MI not from a global contingency table, but from local density approximation. For a d-dimensional continuous variable, if $\epsilon_i$ is the radius to the k-th nearest neighbor of observation $x_i$ and $c_d$ is the unit ball volume in the metric used, then the local density around $x_i$ can be heuristically written as

\begin{equation}
f(x_i)\approx\frac{k}{(N-1)c_d\,\epsilon_i^d}
\end{equation}

Substituting this local density approximation into the definition of differential entropy and replacing the expectation with a sample average yields the Kozachenko--Leonenko estimator \cite{ref13}:

\begin{equation}
h_{KL}(X)=\psi(N)-\psi(k)+\log c_d+\frac{d}{N}\sum_i\log\epsilon_i
\end{equation}

A problem arises when h(X), h(Y), and h(X,Y) are all estimated separately and then substituted into $I(X;Y)=h(X)+h(Y)-h(X,Y)$. In finite samples, the biases of these three estimators do not cancel each other well. Kraskov--Stögbauer--Grassberger (KSG) \cite{ref2} addresses this issue by first finding the radius $\epsilon_i$ to the k-th nearest neighbor in the joint space (X,Y), usually under the \textit{sup norm} or \textit{Chebyshev metric}, then counting the number of neighbors falling within the same radius on the marginals X and Y.

\begin{equation}
\epsilon_i=\operatorname{dist}_k\!\big((x_i,y_i),(X,Y)\big)\ \text{under}\ \lVert\cdot\rVert_\infty
\end{equation}

\begin{equation}
I_{KSG}=\psi(k)+\psi(N)-\frac{1}{N}\sum_i\big[\psi(n_x(i)+1)+\psi(n_y(i)+1)\big]
\end{equation}

The architectural advantage of the KSG (Kraskov--Stögbauer--Grassberger) estimator shines in its ability to suppress dimensional bias. This is achieved by anchoring the computation of the joint radius and the marginal frequency on identical local neighborhood geometry. The method seemingly breathes in tune with the data's topography: contracting automatically when delving into dense observation swirls, and expanding to reach further when traversing sparse information landscapes. This topological elasticity is what determines the strength of kNN. Rather than forcing the entire distribution universe to submit to the rigidity of global histogram compartments, this approach allows the ripples of heavy tails, sporadic regime shifts, and local dependency anomalies to be recorded with high acuity.

Nevertheless, this extreme sensitivity carries its own consequences. In limited sample spaces, the microscopic sensitivity of kNN is highly susceptible to distortion. The method is often trapped, translating \textit{noise}, outliers, or even the friction of liquidity imbalance on the trading floor as intact relational signals. For this reason, mapping with kNN must be read as an extraordinarily sharp yet volatile detection instrument. This characteristic stands as the antithesis of \textit{adaptive binning}, which consistently offers an orderly and cohesive structural arrangement, albeit at the cost of sacrificing analytical resolution depth \cite{ref4}, \cite{ref15}, \cite{ref16}.

\subsection{Community Detection and Clustering Validation}

To eliminate evaluation bias that often arises from disparities in cluster numbers or dimensions, this paper positions \textit{Adjusted Mutual Information} (AMI) as the principal evaluation metric, given its robust ability to neutralize \textit{chance agreement} probability \cite{ref5}.

Furthermore, this evaluation framework refuses to rely solely on the determination of a single community algorithm. Four algorithms---\textit{greedy modularity}, Louvain, Leiden, and Infomap---are employed in parallel to accommodate a diverse spectrum of analytical resolution. On one side, there are approaches that tend to aggressively aggregate massive-scale modules, while on the other side, there are algorithms far more sensitive in dissecting information flow into more precise and fine partitions \cite{ref6}, \cite{ref7}, \cite{ref10}, \cite{ref12}-\cite{ref13}. Through this comprehensive evaluation architecture, the validity of the evaluation of MI, PMFG, and the MI-PMFG synergy can be safeguarded against the bias trap of a single optimization scheme. Rather than focusing on a single viewpoint, this approach enables us to capture the robustness of the same network architecture when dissected and peeled apart by various decoder lenses with differing paradigmatic characteristics.

\section{Data}

Empirically, this experiment executes 24 methodological configurations derived from the intersection of three dependency estimators, a pair of graph filtering filters, and the four community decoders previously described. As the observational foundation, the initial universe was constructed from 150 stocks on the Indonesian exchange. To maintain the purity of the analytical signals, this population was dynamically filtered at each observation window based on strict prerequisites: a minimum data coverage of 60\%, a minimum observation period of 60 days, and an active \textit{listing} status on the \textit{window end} date. This filtering mechanism locks the effective graph dimensions within a very stable range, specifically between 149 and 150 stock nodes (with an average of 149.948).

This entire constellation of methods was then iteratively evaluated across 97 unique \textit{rolling windows}. Each window spanned a length of 300 days with a 25-day shift interval, resulting in a total of 2,328 evaluation observation rows (24 configurations multiplied by 97 windows). Through this architecture, the observation epoch for the \textit{window end} precisely spans from March 23, 2016, to December 12, 2025, derived from the raw data range of January 1, 2015, to December 31, 2025.

Entering the computational implementation phase, the \textit{adaptive binning} model was constructed through a joint histogram framework based on \textit{equal-frequency}. Meanwhile, the MI-kNN estimation was extracted by setting a neighborhood parameter of $k=5$, with its results subsequently smoothed through an iterative \textit{bootstrap} averaging technique ($n_{\textit{bootstrap}}=20$). In weaving its network architecture, the PMFG trajectory was grown \textit{greedily} following the dependency weight hierarchy under strict monitoring of planarity constraints. Conversely, the MST \textit{backbone} was carved purely through a \textit{similarity-to-distance} transformation that absolutely precludes the structural formation of cycles.

\section{Relational Results of the Network Representation}

Overall, the empirical results do not present a single face of the market. The 24 configurations tested instead reveal three layers of network anatomy: a concise linear \textit{backbone}, a more heterogeneous planar extension, and a residual information layer that operates beyond \textit{Gaussian-linear} correlation. Different representations depict the spectral dynamics of stock price movements in Indonesia with distinct characteristics and---certainly---possess their own respective uses and functions.

\begin{figure}[H]
\centering
\includegraphics[width=0.75\textwidth]{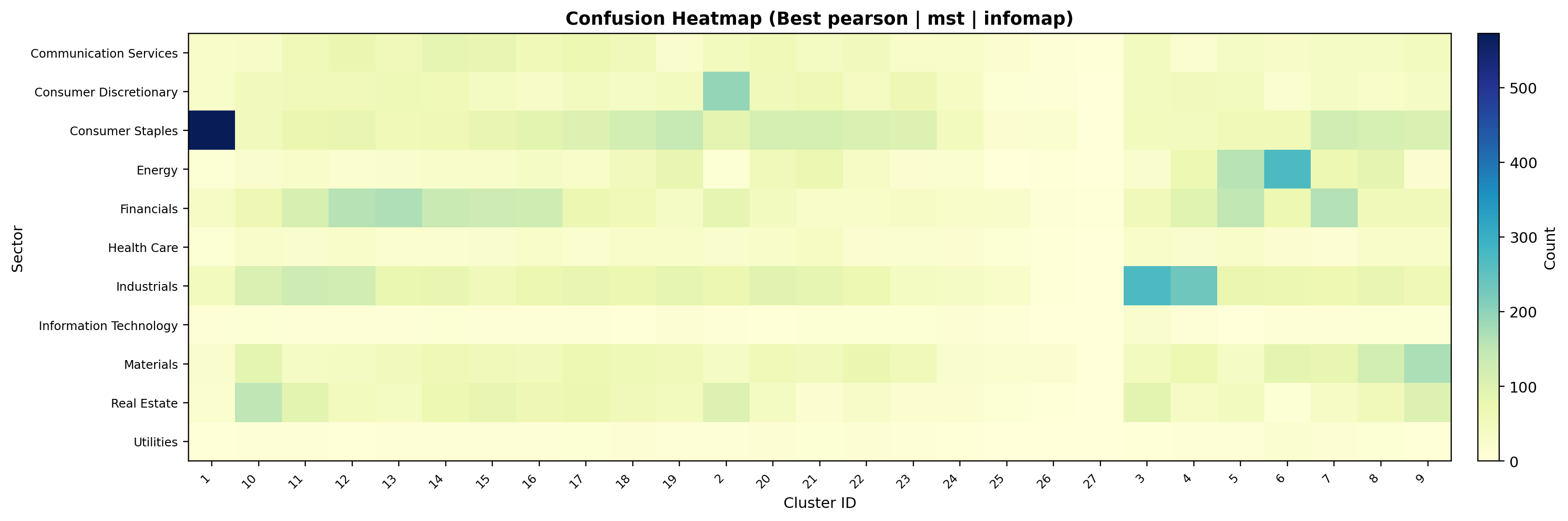}
\caption*{Figure 1 Aggregate confusion heatmap for the best configuration Pearson | MST | Infomap. Sector concentration appears sharper in several dominant clusters.}
\end{figure}

\begin{figure}[H]
\centering
\includegraphics[width=0.75\textwidth]{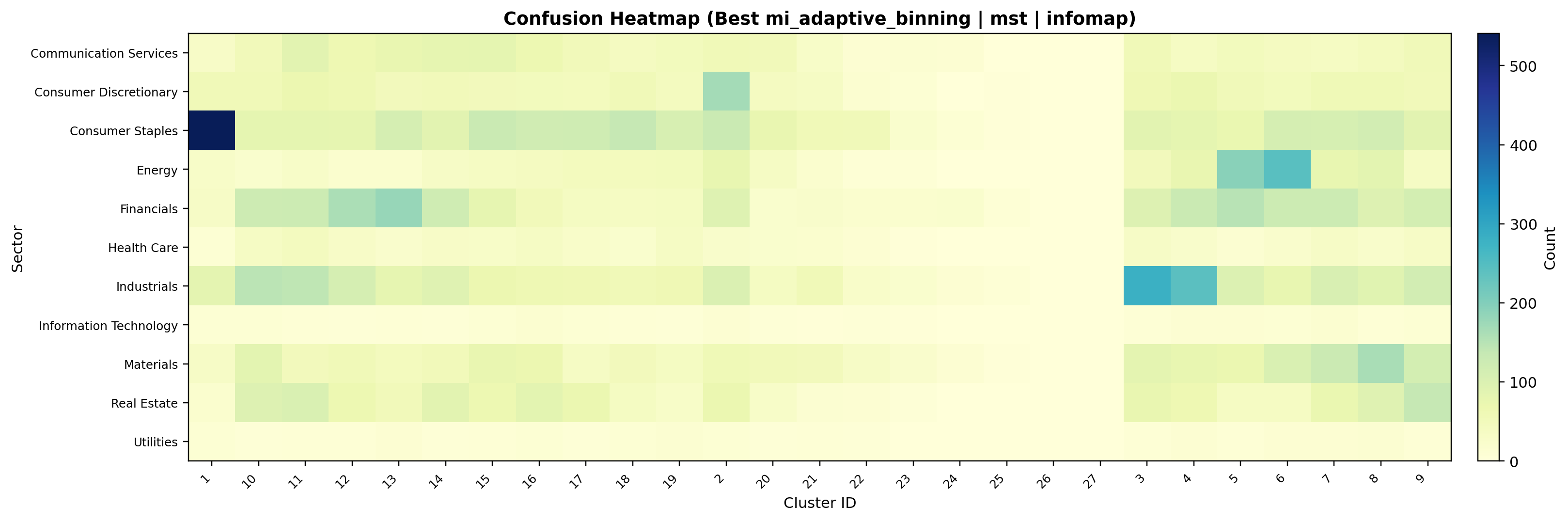}
\caption*{Figure 2 Aggregate confusion heatmap for the best configuration MI adaptive binning | MST | Infomap. Traces of sectoral grouping are visible, with softer inter-community boundaries than under Pearson.}
\end{figure}

The comparison between Figure 1 and Figure 2 displays this difference. The Pearson correlation approach packs the observation mass into a few dominant clusters, so that sectoral cohesion appears more pronounced. MI \textit{adaptive binning} does not destroy that trace but spreads it across several neighboring clusters. In the core region bringing together Financials, Industrials, Materials, and Real Estate, the community boundary lines become softer. This softness is not merely statistical haze. It signals that part of the collective price movement does not fully obey formal sector boundaries; rather, there is relational residue that begins to emerge when the market is read through MI.

The community detection pattern of stocks under the Infomap, MST configuration is observed to fragment the graph into an average of approximately 21.7--24.0 clusters per window, whereas PMFG yields only about 11.0--11.9 clusters. This means that PMFG tends to merge nodes more aggressively, so that the differentiation of stock price movements is more readable than the sector classification, relative to MST. It can be said that the representation under PMFG more clearly displays the intrinsic dynamics of collective price movements with stronger detail.

\begin{figure}[H]
\centering
\includegraphics[width=0.75\textwidth]{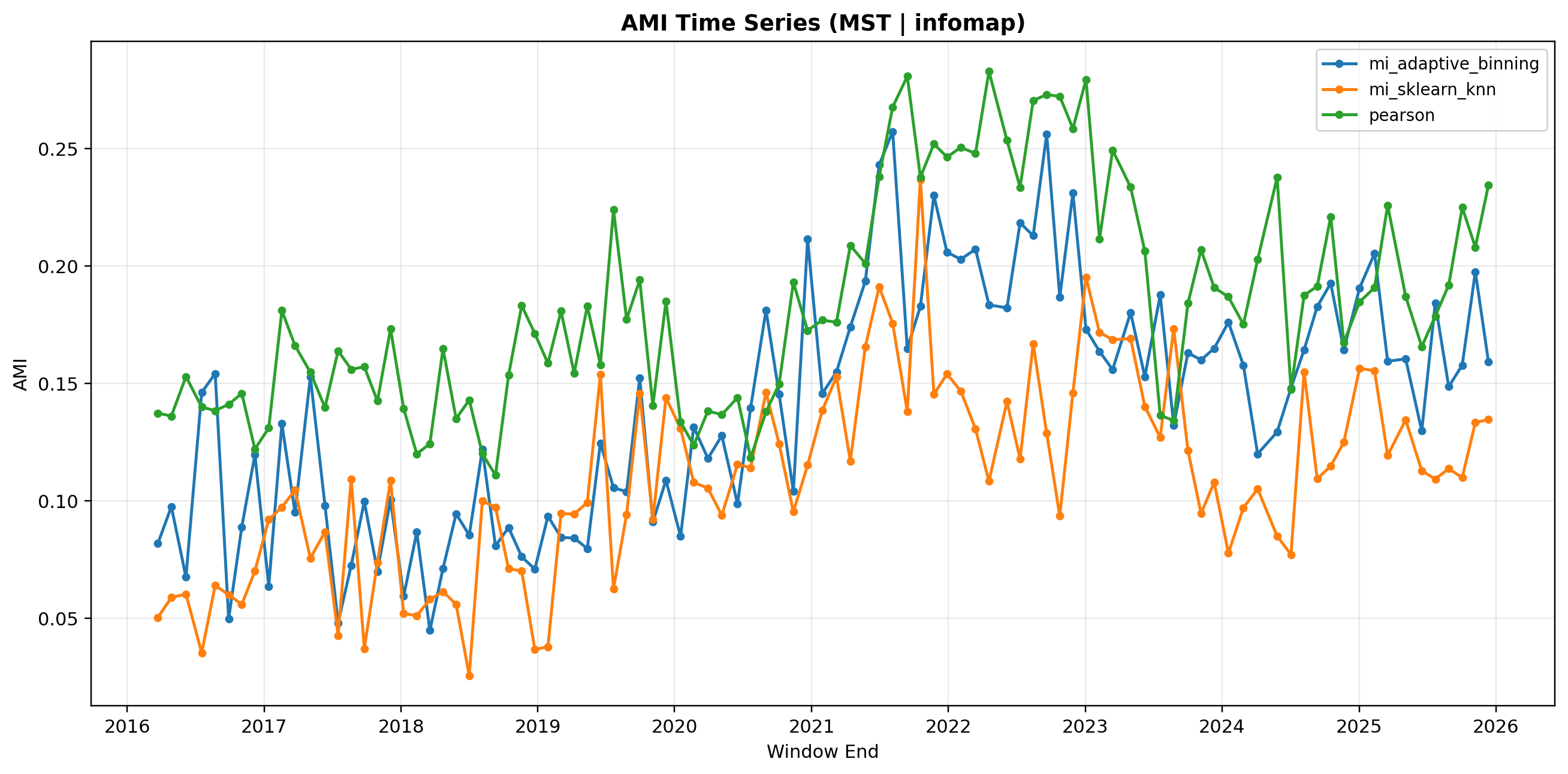}
\caption*{Figure 3 Dynamics of community readability on MST under Infomap.}
\end{figure}

\begin{figure}[H]
\centering
\includegraphics[width=0.75\textwidth]{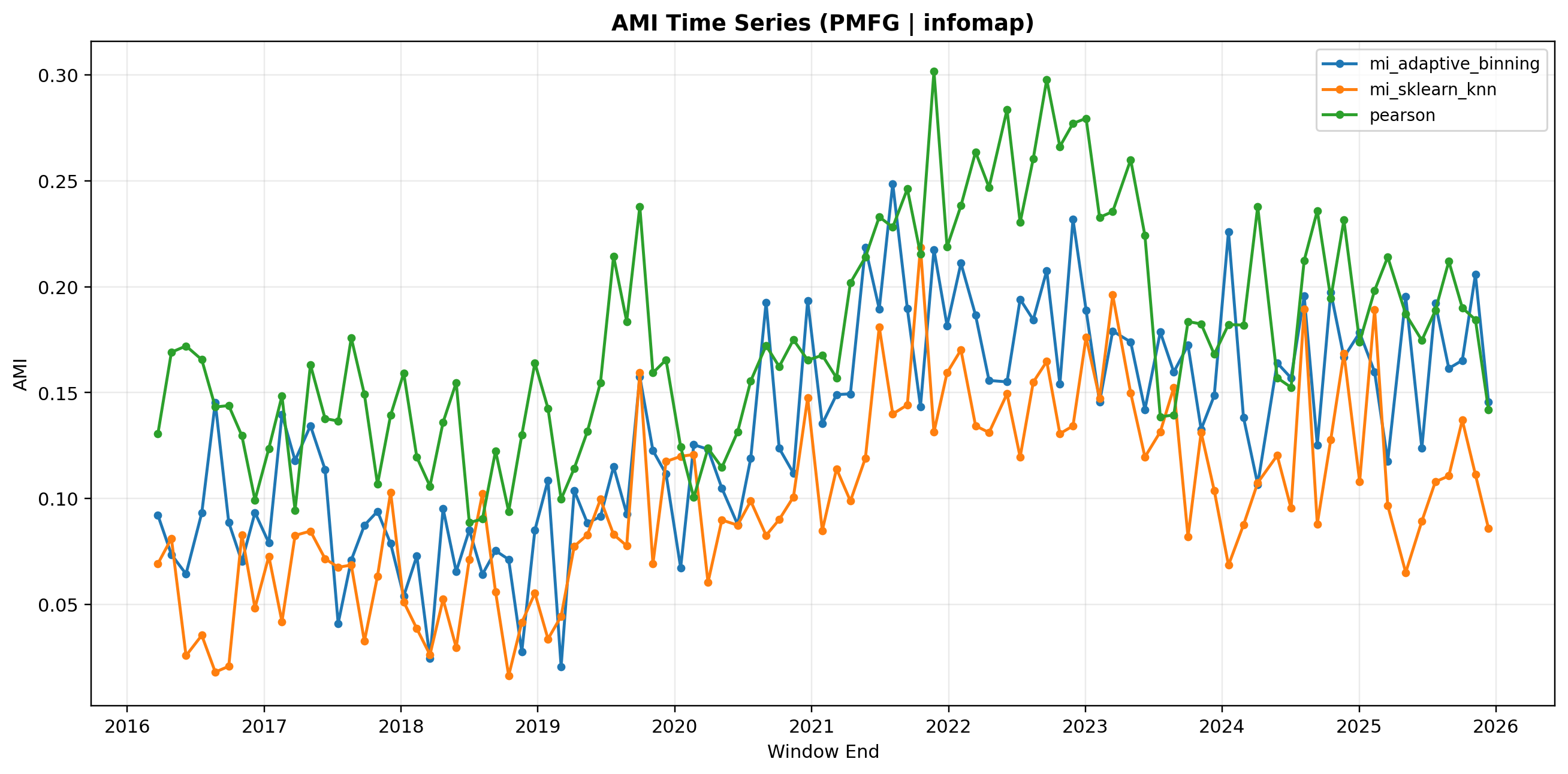}
\caption*{Figure 4 Dynamics of community readability on PMFG under Infomap.}
\end{figure}

The curves in Figure 3 and Figure 4 move in the same direction and both strengthen during the 2021--2022 phase. That phase becomes the moment when market taxonomy is more easily read through the network structure, both on the MST tree and on the PMFG planar graph. However, Pearson still occupies the highest line because it captures the most stable macro pulse: the linear co-movement born of market sentiment, commodities, interest rates, and sectoral cycles. At the same time, the MI line does not lose its meaning. Rather, it signals a more subtle region where information dependency operates as an additional layer beneath the principal correlative cohesion.

Figure 5 brings the reading down to the node level. In the 2022-09-20 \textit{window}, with 150 stocks and the same \textit{layout}, the AMI of Pearson versus MI \textit{adaptive binning} drops from 0.305 to 0.273. This slight decrease is not a sign of vanishing sectoral morphology. The peripheral branches remain similar, but the network core densifies and begins to feature cross-sector \textit{rewiring}. \textit{Adaptive binning} works like a dim lamp that reveals new layers without burning the old map: it preserves the larger shape of the market while adding a finer layer of information dependency.

Figure 6 displays the more restless side of MI-kNN. When the AMI drops from 0.305 to 0.221, sectoral coloring appears more mixed and module boundaries more fragmented. The local neighborhood geometry makes kNN bolder in capturing variations of residual dependency, but that boldness carries consequences: part of the signal becomes more difficult to stabilize as a community. At this point, \textit{adaptive binning} appears more proportional---not as sharp as kNN in pursuing residuals, but better able to maintain structural readability.

\begin{figure}[H]
\centering
\includegraphics[width=0.75\textwidth]{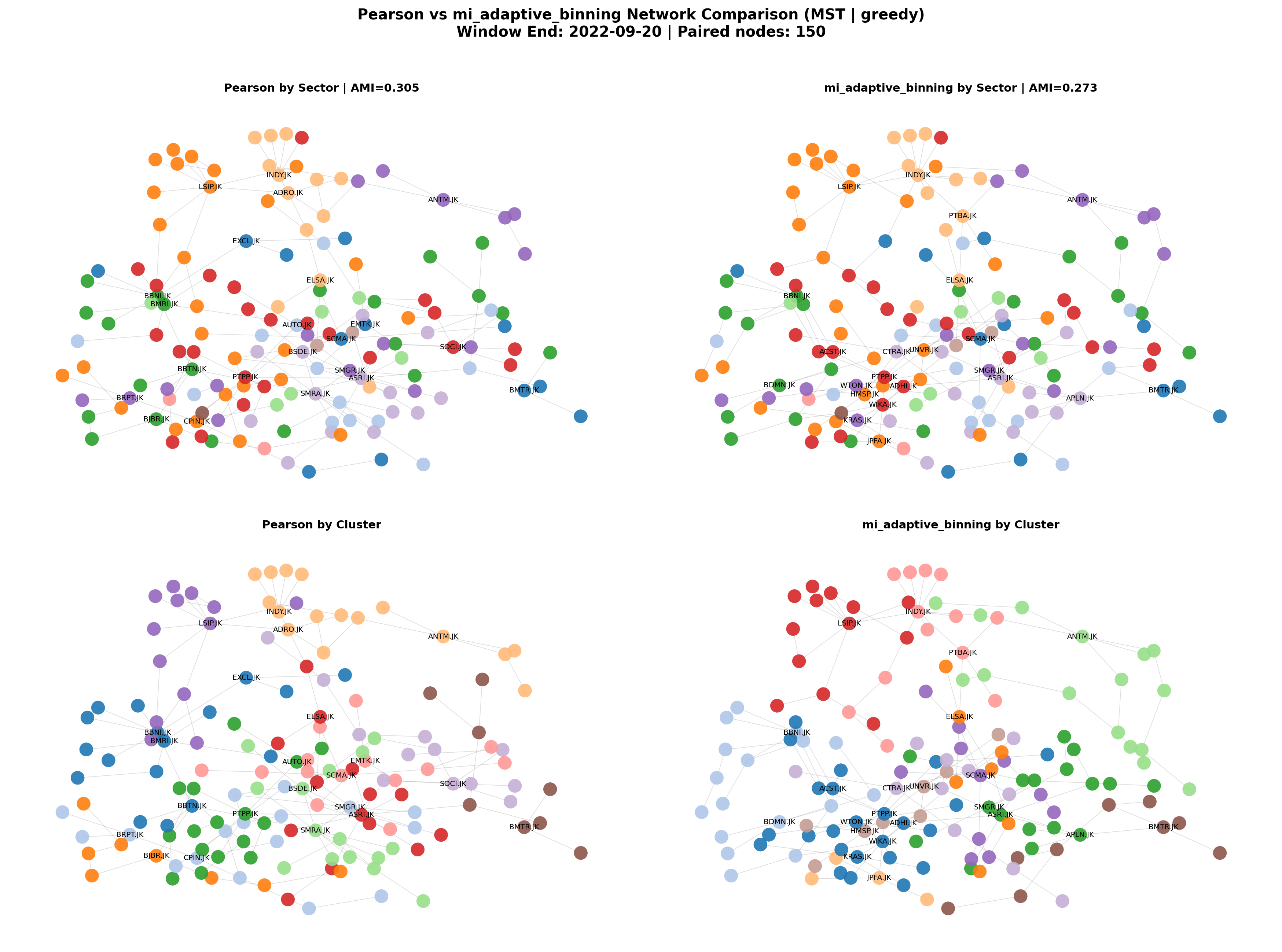}
\caption*{Figure 5 \textit{Node-level} anatomy of Pearson versus MI \textit{adaptive binning} on MST | greedy, \textit{window} 2022-09-20, with identically paired nodes and \textit{layout}.}
\end{figure}

\begin{figure}[H]
\centering
\includegraphics[width=0.75\textwidth]{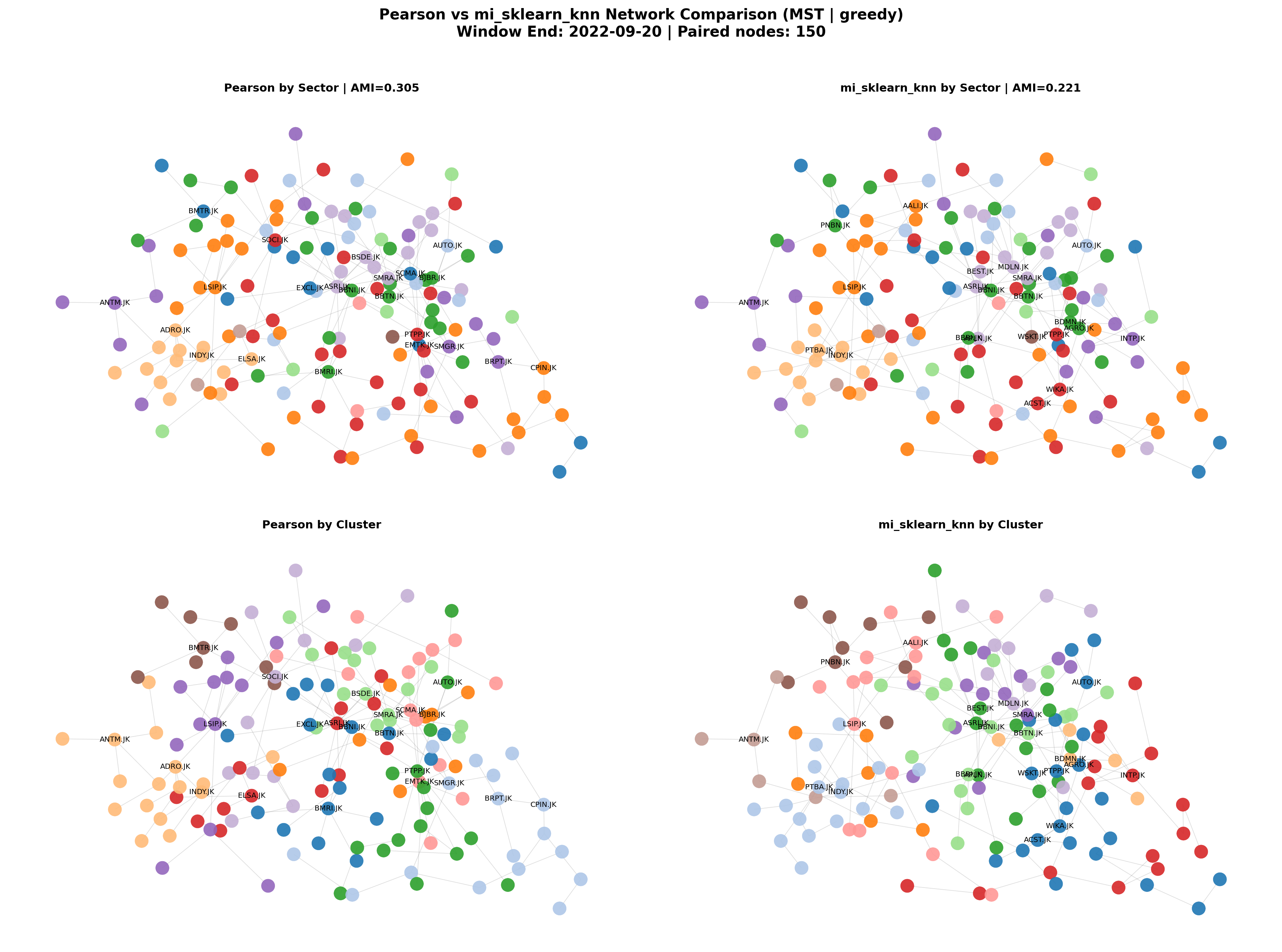}
\caption*{Figure 6 \textit{Node-level} anatomy of Pearson versus MI-kNN on MST | greedy, \textit{window} 2022-09-20, with identically paired nodes and \textit{layout}.}
\end{figure}

From here, the direction of the findings becomes clearer, indicating the representational differences of each method. The Pearson-MST correlative approach provides the macro panorama, PMFG thickens the body of the network, and MI opens the residual layer. Within the \textit{non-linear-planar} family, the MI \textit{adaptive binning} | PMFG | Infomap configuration records an AMI of 0.1331. This value lies below that of Pearson | PMFG | Infomap, which reaches an AMI of 0.1770, but that very difference affirms the function of MI-PMFG: it does not pursue the closest similarity to formal sectors, but rather seeks stock pairs and blocks that still hold shared information after the \textit{Gaussian-linear} \textit{baseline} has been separated.

When \textit{residual information} is translated to the stock-pair level, the door to hidden sub-structures begins to open. Under MI \textit{adaptive binning} | PMFG, the most frequently occurring pairs are concentrated in the energy and mining blocks: INDY-PTBA, HRUM-INDY, ADRO-PTBA, ITMG-PTBA, and DOID-PTBA. Under MI-kNN | PMFG, those blocks persist, and additional links emerge such as ANTM-TINS, ACST-WIKA, and BJBR-BJTM. Economically, these pairs form traces of commodity regimes, construction supply chains, and regional liquidity affinity. These traces indicate diagnostic movements of the price spectrum that transcend the formal sector partition.

\begin{table}[H]
\centering
\caption{Table 1 Stock pairs with high $\Delta I$ that most frequently appear in PMFG residual diagnostics.}
\begin{tabular}{l l l l}
\toprule
\textbf{Estimator} & \textbf{Stock Pair} & \textbf{Mean $\Delta I$} & \textbf{Frequency} \\
\midrule
MI \textit{adaptive binning} & INDY - PTBA & 0.1203 & 120 \\
MI \textit{adaptive binning} & HRUM - INDY & 0.1412 & 83 \\
MI \textit{adaptive binning} & ADRO - PTBA & 0.1311 & 81 \\
MI \textit{adaptive binning} & ITMG - PTBA & 0.1340 & 79 \\
MI \textit{adaptive binning} & DOID - PTBA & 0.0931 & 77 \\
MI-kNN & INDY - PTBA & 0.3469 & 133 \\
MI-kNN & ITMG - PTBA & 0.3426 & 127 \\
MI-kNN & DOID - PTBA & 0.3218 & 106 \\
MI-kNN & ADRO - PTBA & 0.3575 & 95 \\
MI-kNN & ANTM - TINS & 0.3141 & 89 \\
\bottomrule
\end{tabular}
\end{table}

The scale of $\Delta I$ adds an important chapter to the story. MI-kNN repeatedly produces larger $\Delta I$, but that surge does not automatically raise the AMI. This means that a sensor more sensitive to non-linear residuals does not automatically yield communities that are more easily read through market taxonomy. kNN is like a very sharp detector: it captures small ripples, but it is also easily swept along by local noise. \textit{Adaptive binning} offers a calmer compromise between regularization, stability, and structural readability.

Thus, the MI-PMFG approach exhibits a stronger network representation as a diagnostic lens for residual information. MI-PMFG is complementary to the Pearson-MST correlative approach, which provides more of a macro map diagnostically aligned with the commonly known stock sector classifications. Its value lies in the capability to reveal stock pairs or blocks that share extra information beyond linear correlation. This is where the AMI decline under MI-PMFG acquires conceptual meaning: not a blemish in the evaluation, but a sign that the information lens is tracing more hidden relational corridors.

\section{Discussion}

The AMI values help us read which layer of the market is being unveiled by the construction of the price-movement network representation. At the first layer, Pearson-MST captures the most visible co-movements. Stocks within similar sectors often share exposure to market sentiment, commodity prices, interest rates, consumption cycles, and property rotation. For that reason, Pearson-MST communities tend to align with the common sectoral market taxonomy.

The second layer emerges in the network construction with the MI approach. MI does not merely mimic correlation; rather, it seeks the shared information that remains after the \textit{Gaussian-linear} \textit{baseline} has been separated. On a \textit{rolling window} with limited samples, such a search indeed carries a \textit{bias-variance} risk. In a market influenced by commodities, liquidity, and regime shocks, residual relations often live as short episodes rather than as permanent blocks.

PMFG thickens the third layer. Conceptually, the planarity bound provides room for \textit{loops} and small \textit{cliques}, but the empirical results here read this primarily through directly measurable symptoms: fewer \textit{clusters} and larger community sizes. In other words, PMFG is not portraying a "breakdown" of the sectoral structure in the market, but rather expanding the network anatomy so that interlocking market blocks can emerge together. When communities become heterogeneous, we are viewing the market through a thicker body than the MST tree, rather than through a thin sectoral map.

From the graph-construction side, that difference is reasonable. MST selects N-1 edges and builds the most economical path to connect all stocks. PMFG continues to add edges as long as planarity is preserved, up to the bound of 3N-6. These additional edges make the network richer, but also render it more sensitive to small changes in the weight ordering. A single edge rising or falling near the threshold can alter the planar relations formed in the next \textit{window}. For this reason, link stability must be read as a separate matter, not merely a derivative of the magnitude of correlation \cite{ref25}.

The comparison between \textit{adaptive binning} and MI-kNN displays two different methodological philosophies. \textit{Adaptive binning} calms the data with \textit{quantile-based coarse-graining}, dampening the sparsity of the distribution tails and restraining excessively wild \textit{rewiring} \cite{ref14}, \cite{ref16}. MI-kNN moves in the opposite direction: delving into the geometry of the local neighborhood and being more sensitive to \textit{tail dependence}, \textit{switching regime}, and \textit{non-Gaussianity} \cite{ref4}, \cite{ref15}, \cite{ref16}. For this reason, kNN can produce larger $\Delta I$, but its topology vibrates more easily. In the context of the Indonesia Stock Exchange, \textit{adaptive binning} becomes a middle ground more aligned with the need to read structure, rather than merely pursuing the most extreme residuals.

This functional division of roles is the crux of the discussion. Pearson-MST is most useful as a macro panorama: providing a clean linear backbone that is easily linked to common market taxonomy. Pearson-PMFG is useful for reading the planar expansion: preserving partial sectoral readability while beginning to display thicker local cohesion. MI-PMFG, especially through \textit{adaptive binning} and Infomap, works as a diagnostic lens: tracing hidden sub-structures, commodity regimes, and residual relations that are not fully visible in linear correlation. Thus, the score differences between methods are not the end of the debate, but rather an indication of the market layer we are reading.

\section{Conclusion}

Various typologies of reading the representation of price movements in the stock market as a network have been demonstrated. First, the network using the Pearson correlation | MST | Infomap approach produces the purest linear \textit{backbone}: showing that its communities resonate most easily with the formal economic divisions. Second, PMFG expands that \textit{backbone} into a thicker planar anatomy. Its communities are larger and more heterogeneous, so that the score against the market taxonomy decreases, but the relational space being read also becomes wider. Third, MI \textit{adaptive binning} | PMFG | Infomap becomes the most proportional MI-PMFG configuration for reading the residual information without losing all structural readability.

MST provides a concise map for viewing the macro arrangement of the market. PMFG reveals the planar layers not visible on the tree. MI adds an informational dimension that opens up stock pairs and blocks with high $\Delta I$, especially in sectors such as energy, mining, construction, and regional liquidity affinity. Thus, the AMI decline under PMFG or MI can be read as a sign of a lens shift from sectoral readability toward structural-residual reading.

Functionally, the use of these three price-movement network approaches can be clearly differentiated. When a macro depiction of collective price movement is required, Pearson-MST provides the most concise framework. When attention is directed to the thickness of local structure and the planar relationships between market blocks, Pearson-PMFG becomes a sensible choice. When the question shifts to non-linear dependency, \textit{Gaussian-linear} residuals, and hidden sub-structures, MI-PMFG provides the most valuable analytical lens. The representation of collective stock-price movements provides a medium for reading the spectral dynamics of the market more completely: from the correlational backbone, through planar expansion, to the hidden residual-information corridor lying behind the collective movement of prices.

\end{document}